\documentclass[useAMS,usenatbib]{mn2e}

\title[
H$_2$ Excitation in Bipolar Planetary Nebulae
]{
The Excitation Mechanism of H$_2$ in Bipolar Planetary Nebulae
}
\author[Marquez-Lugo, R. A.; Guerrero, M. A.;  Ramos-Larios, G. and, Miranda, L. F. ]{
R.A.\ Marquez-Lugo,$^{1}$\thanks{E-mail: amarquez@iaa.es} 
M.A.\ Guerrero,$^{1}$ 
G.\ Ramos-Larios$^{2}$ and 
L.F.\ Miranda$^{1}$
\\
$^{1}$Instituto de Astrof\'{\i}sica de Andaluc\'{\i}a IAA, CSIC, 
Glorieta de la Astronom\'{\i}a s/n, E-18008 Granada, Spain \\
$^{2}$Instituto de Astronom\'ia y Meteorolog\'ia, CUCEI, Universidad de Guadalajara,
Av. Vallarta No. 2602, Col. Arcos Vallarta, \\
44130 Guadalajara, Jalisco,
Mexico\\
}
\usepackage{subfig}
\usepackage{graphicx}
\usepackage{booktabs}
\usepackage{tablefootnote}

\begin{document}

\date{today}

\pagerange{\pageref{firstpage}--\pageref{lastpage}} \pubyear{2014}

\maketitle

\label{firstpage}

\begin{abstract}
We present near-IR $K$-band intermediate-dispersion spatially-resolved 
spectroscopic observations of a limited sample of bipolar planetary 
nebulae (PNe).  
The spectra have been used to determine the excitation mechanism 
of the H$_2$ molecule using standard line ratios diagnostics.  
The H$_2$ molecule is predominantly shock-excited in bipolar PNe with broad 
equatorial rings, whereas bipolar PNe with narrow equatorial waists present 
either UV excitation at their cores (e.g., Hb\,12) or shock-excitation at 
their bipolar lobes (e.g., M\,1-92).  
The shock-excitation among bipolar PNe with ring 
is found to be correlated with emission in the H$_2$ 1-0 S(1) line 
brighter than Br$\gamma$. 
We have extended this investigation to other PNe with available near-IR 
spectroscopic observations.  
This confirms that bipolar PNe with equatorial rings are in average 
brighter in H$_2$ 
than in Br$\gamma$ 
and show dominant shock excitation.  
\end{abstract}


\section{Introduction}

Planetary Nebulae (PNe) represent the short-lived transition from red 
gianthood to white dwarfdom for low- and intermediate-mass (1-8~M$_{\odot}$) 
stars. 
These stars eject most of their envelopes at the end of the 
asymptotic giant branch (AGB). 
The ejecta forms a circumstellar shell which is subsequently 
ionized by the central star (CSPN) and becomes a PN.  
While the optical emission from PNe is dominated by emission lines from 
ionized gas, molecular material may still survive wherever the optical 
depth to ionizing photons is high, either in high-density clumps or in 
photo-dissociation regions \citep[PDR,][]{1993IAUS..155..155T}.

Near-infrared imaging of the H$_2$ 1-0 S(1) $\lambda$2.122 $\mu$m 
emission line has been used to detect molecular material at high 
spatial resolution and sensitivity.
Since the first detection of H$_2$ emission in NGC\,7027 
\citep{1976ApJ...209..793T}, the near-IR H$_2$ 1-0 S(1) 
line has been detected in over 100 PNe and proto-PNe 
\citep[][and references therein]{1996ApJ...462..777K, 1999ApJS..124..195H, 2000ApJS..127..125G, 2002A&A...387..955G}.  
These observations show that H$_2$ emission is predominantly detected 
in the equatorial regions of bipolar PNe (hereafter BPNe), as self- and 
dust-shielding of UV photons at PDRs in their dense equatorial regions 
can provide a safe haven for molecules.  
Although BPNe are the brightest H$_2$-emitters, the presence of molecular 
hydrogen is not exclusive to BPNe; sensitive observations have proven that 
H$_2$ emission can also be detected in PNe with ellipsoidal or barrel-like 
morphologies \citep{2013MNRAS.429..973M}.  
The high-spatial resolution of these observations unveil that the H$_2$ 
emission from the equatorial ring of BPNe arises from knots and clumps 
embedded within the ionized material of the ring rather than from a PDR 
\citep{Manchado15}.  
This is also in concordance with the detection of H$_2$ emission in bipolar 
lobes of BPNe \citep[e.g.,][]{RLGM2008}, which points at the 
presence of molecular material in an environment different from 
a PDR.  

\begin{figure*}
  \centering
  \includegraphics[trim = 5mm 5mm 175mm 260mm, clip, width=0.9\textwidth]{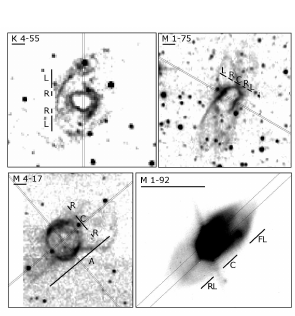}
  \caption{\label{rendijas1}
Images in the near-IR H$_2$ filter of the H$_2$-bright BPNe in our sample.  
The slit positions of the near-IR spectroscopic observations and the 
extent of the spatial apertures used in the extraction of one-dimensional 
spectra are labeled (see Section 2.1). 
In all figures, North is up, East left.  
The horizontal bar below the name of each PN corresponds to 5\arcsec.  
Note that the linear feature along PA$\sim$10$^\circ$ in the image of 
M\,1-92 is an artifact.  }
\end{figure*}

Morphologically, the class of BPNe is not homogeneous, with two distinct 
sub-classes \citep{1996iacm.book.....M}: 
BPNe with a compact core or narrow waist (W-BPNe), and  
BPNe with a broad ring structure in the waist (R-BPNe). 
It is unclear whether these two sub-classes of BPNe descend from progenitors 
in the same mass range, but observed at different evolutionary stages, or 
they proceed from progenitors of different mass ranges and follow different 
evolutionary paths  \citep{2000ApJS..127..125G}.  
R-BPNe exhibit the brightest H$_2$ emission and 
the highest H$_2$/Br$\gamma$ flux ratio 
\citep{1988MNRAS.235..533W, 2000ApJS..127..125G}.  
It is tempting to correlate the H$_2$ emission with the amount of molecular 
mass and conclude that R-BPNe have larger reservoirs of molecular material 
and proceed from more massive progenitors.  
The intensity of the H$_2$ emission depends strongly on the excitation 
mechanism, however.  
In the particular case of BPNe, the typical shock velocities imply 
higher H$_2$ emission line intensities than those expected for simple 
UV-excitation \citep{Burton92}.  
Therefore, imaging in the 2.122 $\mu$m H$_2$ line alone cannot be used to 
determine the total molecular mass to assess whether the two sub-classes 
of BPNe are related to different evolutionary stages of the same class of 
object, or to distinct classes of objects. 

To determine the molecular content of the two different sub-classes of 
BPNe, first we have to establish the excitation conditions of H$_2$.  
We have then selected the sample of BPNe 
listed in Table~\ref{obs_spec} 
and obtained moderate-resolution near-infrared 
$K$-band spectroscopy to diagnose whether the 
H$_2$ line ratios are consistent with shock or 
UV excitation \citep{1987ApJ...322..412B}.  
The spectroscopic observations are described in Sect.\ 2, the spectra 
and line intensity measurements are presented in Sect.\ 3, and the 
investigation of the H$_2$ excitation mechanism in these PNe is shown 
in Sect.\ 4.  
The results are discussed in Sect.\ 5 and the main conclusions 
summarized in Sect.\ 6.  

\section[]{Observations}

Near-IR $K$-band spectrophotometric observations of the seven BPNe 
listed in Table~\ref{obs_spec} were obtained on July 2-4, 2004 using 
the 3.5m Telescopio Nazionale Galileo (TNG) and the Near Infrared 
Camera Spectrometer (NICS) in its spectroscopic mode. 
A HAWAII 1024$\times$1024 detector was used, resulting in a spectral 
dispersion of 4.3 \AA~pixel$^{-1}$ (4.3$\times$10$^{-4}$ $\mu$m~pixel$^{-1}$) 
and a wavelength coverage from 1.925 $\mu$m to 2.368 $\mu$m.  
The seeing during the observations varied from 0\farcs7 up to 1\farcs2. 

For the slit width of 0\farcs75 used during the observations, the 
spectral resolution was $\simeq$13{\AA} for a resolution power 
$R\simeq$1630.  
The slit position angles (PAs) and total integration times are given in 
columns 4 and 5 of Table~\ref{obs_spec}, respectively.  
The spectra were reduced using standard {\sc IRAF}\footnote{
{\sc iraf}, the Image Reduction and Analysis Facility, is distributed 
by the National Optical Astronomy Observatory, which is operated by the 
Association of Universities for Research in Astronomy under cooperative 
agreement with the National Science Foundation.} 
V2.14.1 routines.  
The wavelength calibration was obtained using telluric sky lines.  
The flux calibration was carried out using the standard IR stars of spectral 
type A--A0 SAO 15832, SAO 48300, and SAO 72320 \citep{1998AJ....115.2594H}. 
These same stars were used to correct the nebular spectra 
from telluric absorptions.

\begin{figure}
  \centering
\includegraphics[trim = 4mm 5mm 187mm 250mm, clip, width=0.45\textwidth]{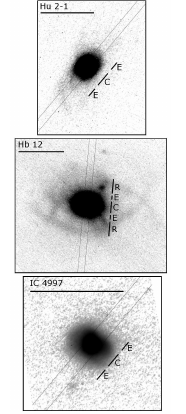}
  \caption{\label{rendijas2}
Images in the near-IR H$_2$ filter of the H$_2$-weak BPNe in our sample.  
The slit positions of the near-IR spectroscopic observations and the 
extent of the spatial apertures used in the extraction of one-dimensional 
spectra are labeled. 
In all figures, North is up, East left.  
The horizontal bar below the name of each PN corresponds to 5\arcsec. 
} 
\end{figure}

The slits go across the nebular center in all cases along directions of
special interest.  
Figures~\ref{rendijas1} and \ref{rendijas2} show the slit positions 
on near-IR images of the PNe.  
Details of the imaging observations are given in Table~\ref{obs_img}.  
In the cases of K\,4-55, M\,1-75, and M\,4-17, Figure~\ref{rendijas1} 
shows the Calar Alto (CAHA) 2.2 m telescope H$_2$ images presented by 
\citet{2000ApJS..127..125G}, where more details on the observations 
and data reduction are given.  
For the other four PNe (Figures~\ref{rendijas1} and ~\ref{rendijas2}), 
narrow-band near-IR images were obtained at the TNG with NICS in its 
imaging mode using the Short Field camera in combination with the 
adaptive optics module.  
The detector was a 1024$\times$1024 HgCdTe array with pixel size of 
0\farcs04 and field of view of 0\farcm7$\times$0\farcm7.  
Images of Hu\,2-1, M\,1-92, and IC\,4497 were obtained on 2003 September 
17-18, while Hb\,12 was observed on 2004 September 17. 
The filter details and exposure times are given in Table~\ref{obs_img}.  
The spatial resolution of the images is between 0\farcs4 and 0\farcs6.  
The images were reduced following standard procedures. 
Figures~\ref{rendijas1} and ~\ref{rendijas2} show the H$_2$ images 
of Hu\,2-1, M\,1-92, IC\,4497, and Hb\,12. 
Their Br$\gamma$ and K$_{cont}$ images (not shown here) are 
very similar to the H$_2$ ones at the angular resolution achieved by 
these images, showing dominant core emission in all of them and weaker 
extended emission in the cases of Hb\,12 and M\,1-92.  
An attempt to substract the continuum contribution from the H$_2$ and 
Br$\gamma$ line images did not yield satisfactory results given the 
bright continuum emission at the core of these nebulae.

\begin{table}
 \centering
 \label{1}
  \caption{\label{obs_spec}Details of the spectral observations.}
  \begin{tabular}{@{}lccrc@{}}
  \hline
Source    & $\alpha$ (J2000) & $\delta$ (J2000) & P.A. & Exposure time \\
          &                  &                  & ($^\circ$)~~ &    (s)  \\
 \hline
 Hb\,12    & 23 26 14.8 & $+$58 10 55 & $-$5 & 200 \\
 Hu\,2-1   & 18 49 47.6 & $+$20 50 39 &  320 & 600 \\
 IC\,4997  & 20 20 08.7 & $+$16 43 54 &  140 & 600 \\
 K\,4-55   & 20 45 10.0 & $+$44 39 16 &    0 & 600 \\
 M\,1-75   & 20 04 44.1 & $+$31 27 24 &   60 & 450 \\
 M\,1-92   & 23 26 14.8 & $+$58 10 55 &  313 & 120 \\
 M\,4-17   & 20 09 01.9 & $+$43 43 44 &   40 & 600 \\
           &            &             &  130 & 600 \\
 \hline
\end{tabular}
\end{table}

\begin{table*}
 \centering
 \label{1}
  \caption{\label{obs_img}Details of the imaging observations.}
  \begin{tabular}{@{}lcccccr@{}}
  \hline
Source    &  Telescope & Instrument & Filter & $\lambda_c$ & $\Delta\lambda$ & Exposure time \\
          &            &            &        &   ($\mu$m)  &  ($\mu$m)       &   (s)~~~~~~  \\
 \hline
Hb\,12    &   TNG      &   NICS     & H$_2$  &      2.122  &     0.032       &   720~~~~~~   \\
Hu\,2-1   &   TNG      &   NICS     & H$_2$  &      2.122  &     0.032       &  1620~~~~~~   \\
IC\,4997  &   TNG      &   NICS     & H$_2$  &      2.122  &     0.032       &  1800~~~~~~   \\
K\,4-55   &  2.2m CAHA &  MAGIC     & H$_2$  &      2.122  &     0.021       &   100~~~~~~   \\
M\,1-75   &  2.2m CAHA &  MAGIC     & H$_2$  &      2.122  &     0.021       &   100~~~~~~   \\
M\,1-92   &   TNG      &   NICS     & H$_2$  &      2.122  &     0.032       &   600~~~~~~   \\
M\,4-17   &  2.2m CAHA &  MAGIC     & H$_2$  &      2.122  &     0.021       &   100~~~~~~   \\
 \hline
\end{tabular}
\end{table*}

\begin{table}
\centering
\begin{minipage}{185mm}
\label{1}
 \caption{\label{H2}Br$\gamma$ normalized fluxes of H$_2$-bright BPNe.}
\rotatebox{90}{
\begin{tabular}{@{}lcccccccccccc@{}}
 \hline
 Line ID     &Wavelength & \multicolumn{2}{c}{K\,4-55}        &  
\multicolumn{3}{c}{M\,1-75} & \multicolumn{3}{c}{M\,4-17} &
\multicolumn{3}{c}{M\,1-92} \\
  &  ($\mu$m) & \multicolumn{2}{c}{P.A. 0$^\circ$} & 
\multicolumn{3}{c}{P.A. 60$^\circ$} & \multicolumn{1}{c}{P.A.130$^\circ$}&
\multicolumn{2}{c}{P.A.40$^\circ$} & \multicolumn{3}{c}{P.A.313$^\circ$}
\\
\cmidrule(lr){3-4}\cmidrule(lr){5-7}\cmidrule(lr){8-10}\cmidrule(lr){11-13}
   &&Lobes & Ring & Lobes & Ring &Center & All & Ring & Center & F. lobe
& R. lobe & Center\\
\hline
H$_2$ 1-0 S(3) & 1.9570 &  400$\pm$50 & 217$\pm$18  & 480$\pm$50 &  140$\pm$10  & $\cdots$     & 196$\pm$27   & 345$\pm$31 &   82$\pm$4   &   49$\pm$11  & 838$\pm$34 & $\cdots$      \\
H$_2$ 1-0 S(2) & 2.0334 &  510$\pm$50 & 185$\pm$17  & 211$\pm$33 &   48$\pm$6   & $\cdots$     & 139$\pm$14   & 247$\pm$26 &   91$\pm$4   &    56$\pm$4  & 297$\pm$20 & $\cdots$      \\
He I           & 2.0580 &  $\cdots$   &  39$\pm$8   &  96$\pm$22 &   34$\pm$5   & $\cdots$     &   65$\pm$7   &  32$\pm$10 & 48.2$\pm$2.3 & $\cdots$     & $\cdots$   & 300$\pm$4     \\
H$_2$ 2-1 S(3) & 2.0732 &  104$\pm$24 &  33$\pm$7   &  27$\pm$12 &  4.9$\pm$1.8 & $\cdots$     & 21.5$\pm$3.1 &  47$\pm$12 & 14.5$\pm$1.2 & 17.5$\pm$2.3 &  79$\pm$11 & $\cdots$      \\
H$_2$ 1-0 S(1) & 2.1218 & 1370$\pm$90 & 452$\pm$26  & 830$\pm$70 &  176$\pm$11  & 47.7$\pm$2.8 & 307$\pm$31   & 530$\pm$40 & 210$\pm$9    & 153$\pm$7    & 723$\pm$32 & $\cdots$      \\
H$_2$ 2-1 S(2) & 2.1542 &  110$\pm$22 &  16$\pm$5   &  36$\pm$14 &  6.7$\pm$2.1 & $\cdots$     &  52$\pm$5    &  53$\pm$12 & 17.2$\pm$1.3 & 18.9$\pm$2.4 &  58$\pm$9  & $\cdots$      \\
Br$\gamma$     & 2.1658 &  100$\pm$22 & 100$\pm$13  & 100$\pm$23 &  100$\pm$8   & 100$\pm$5    & 100$\pm$10   & 100$\pm$17 & 100$\pm$4    & 100$\pm$6    & 100$\pm$12 & 100.0$\pm$2.0 \\
He II          & 2.1892 &  $\cdots$   &  28$\pm$7   & $\cdots$   &   30$\pm$5   & 61.9$\pm$2.5 & $\cdots$     & $\cdots$   & $\cdots$     & $\cdots$     & $\cdots$   & $\cdots$      \\
H$_2$ 1-0 S(0) & 2.2234 & 310$\pm$40  &  107$\pm$13 & 212$\pm$31 &   45$\pm$6   & $\cdots$     &  98$\pm$10   & 120$\pm$18 &  88$\pm$4    &   44$\pm$4   & 213$\pm$17 & $\cdots$      \\
H$_2$ 2-1 S(1) & 2.2426 & 152$\pm$29  &   25$\pm$6  &  67$\pm$19 & 14.3$\pm$3.1 & $\cdots$     &  47$\pm$5    &  61$\pm$13 & 27.2$\pm$1.8 & 30.4$\pm$3.1 & 105$\pm$12 & $\cdots$      \\
& &  &  &  &  &  & &  &  \\
F(Br$\gamma$)$^a$ & 2.1658 & 8.1 & 33 & 9.8 & 74 & 11 &  120 & 25 & 52 & 160 & 36 & 1300  \\
\hline
\multicolumn{13}{l}{\footnotesize{
(a) Flux of the  Br$\gamma$ line in units of 10$^{-16}$ erg~cm$^{-2}$~s$^{-1}$. }}
\end{tabular}
}
\end{minipage}
\end{table}

\begin{table*}
\centering
 \caption{\label{HI}Br$\gamma$ normalized fluxes of H$_2$-weak BPNe.}
 \begin{tabular}{@{}llccccccc@{}}
 \hline
\multicolumn{1}{l}{Line ID}    & 
\multicolumn{1}{c}{Wavelength} & 
\multicolumn{2}{c}{IC\,4997}   &
\multicolumn{2}{c}{Hu\,2-1}    & 
\multicolumn{3}{c}{Hb\,12}     \\
   &  
\multicolumn{1}{c}{($\mu$m)}        & 
\multicolumn{2}{c}{P.A. 140$^\circ$} &
\multicolumn{2}{c}{P.A. 320$^\circ$} &
\multicolumn{3}{c}{P.A.--5$^\circ$}  \\
\cmidrule(lr){3-4}\cmidrule(lr){5-6}\cmidrule(lr){7-9}
   &&Envel. & Center & Envel. & Center & Ring & Envelope & Core  \\
\hline
He I           & 1.9541 & $\cdots$ & $\cdots$ & $\cdots$ & 16.2$\pm$0.6 & $\cdots$ & $\cdots$ & $\cdots$ \\
H$_2$ 1-0 S(3) & 1.9570 & $\cdots$ & $\cdots$ & $\cdots$ & $\cdots$     & $\cdots$ & $\cdots$ & 11.8$\pm$0.3  \\
H$_2$ 1-0 S(2) & 2.0334 & $\cdots$ & $\cdots$ & $\cdots$ & $\cdots$     & $\cdots$ & 10.1$\pm$1.6 & $\cdots$   \\
H$_2$ 8-6 S(3) & 2.0425 & $\cdots$ & 1.7$\pm$0.1 & $\cdots$ & $\cdots$ & $\cdots$ & 10.6$\pm$1.6 & 2.3$\pm$0.1 \\
He I           & 2.0580 & 25.4$\pm$1.3 & 33.8$\pm$1.4 & 68$\pm$5 & 80.1$\pm$1.4 & 103$\pm$7 & 95$\pm$5 & 107.7$\pm$0.5  \\
H$_2$ 2-1 S(3) & 2.0732 & $\cdots$  & $\cdots$  & $\cdots$  & $\cdots$  & $\cdots$  & 9.0$\pm$1.5 & $\cdots$  \\
He I           & 2.1124 & $\cdots$  & 8.6$\pm$0.4 & 7.8$\pm$1.7 & 7.7$\pm$0.4 & 12.3$\pm$2.5 & $\cdots$  & 9.2$\pm$0.2 \\
H$_2$ 1-0 S(1) & 2.1218 & $\cdots$  & 0.6$\pm$0.1 & $\cdots$  & $\cdots$  & 14.8$\pm$2.8 & 25.1$\pm$2.5 & $\cdots$   \\
Fe III         & 2.1460 & $\cdots$  & $\cdots$  & $\cdots$  & $\cdots$  & $\cdots$  & 5.7$\pm$1.2 & $\cdots$  \\
H$_2$ 2-1 S(2) & 2.1542 & $\cdots$  & $\cdots$  & $\cdots$  & $\cdots$  & $\cdots$  & 6.0$\pm$1.2 & $\cdots$   \\
He II          & 2.1614 & $\cdots$  & 4.2$\pm$0.2 & 6.3$\pm$1.5 & 4.9$\pm$0.3 & $\cdots$  & $\cdots$  & 4.1$\pm$0.1 \\
Br$\gamma$     & 2.1658 & 100$\pm$4 & 100$\pm$4 & 100$\pm$6 & 100.0$\pm$1.5 & 100$\pm$7 & 100$\pm$5 & 100.0$\pm$0.5 \\
He I           & 2.1815 & $\cdots$  & 0.9$\pm$0.1 & $\cdots$  & $\cdots$  & $\cdots$  & $\cdots$  & 1.5$\pm$0.1  \\
Kr III         & 2.1993 & $\cdots$  & $\cdots$  & $\cdots$  & 2.7$\pm$0.3 & $\cdots$  & $\cdots$  & $\cdots$   \\
H$_2$ 3-2 S(3) & 2.2014 & $\cdots$  & $\cdots$  & $\cdots$  & $\cdots$  & $\cdots$  & 7.7$\pm$1.4 & $\cdots$  \\
Fe III         & 2.2184 & $\cdots$  & 1.4$\pm$0.1 & $\cdots$  & 1.2$\pm$0.2 & $\cdots$  & 17.6$\pm$2.1 & 1.7$\pm$0.1 \\
H$_2$ 1-0 S(0) & 2.2234 & $\cdots$  & $\cdots$  & $\cdots$  & $\cdots$  & 14.8$\pm$2.7 & 15.7$\pm$2.0 &  $\cdots$ \\
H$_2$ 2-1 S(1) & 2.2426 & $\cdots$  & 0.5$\pm$0.1 & $\cdots$  & $\cdots$  & 18.9$\pm$3.1 & 12.1$\pm$1.7 & $\cdots$   \\
               &        &           &             &           &           &              &              &   \\
F(Br$\gamma$)$^a$  & 2.1658 & 63 & 6100 & 140 & 2200 & 100 & 210 & 19000  \\
\hline
\multicolumn{9}{l}{\footnotesize{
(a) Flux of the  Br$\gamma$ line in units of 10$^{-16}$ erg~cm$^{-2}$~s$^{-1}$. }}
\end{tabular}
\end{table*}

\begin{figure*}
  \centering
  \subfloat[K 4-55]{
  \includegraphics[trim = 10mm 110mm 10mm 15mm, clip, width=0.45\textwidth]{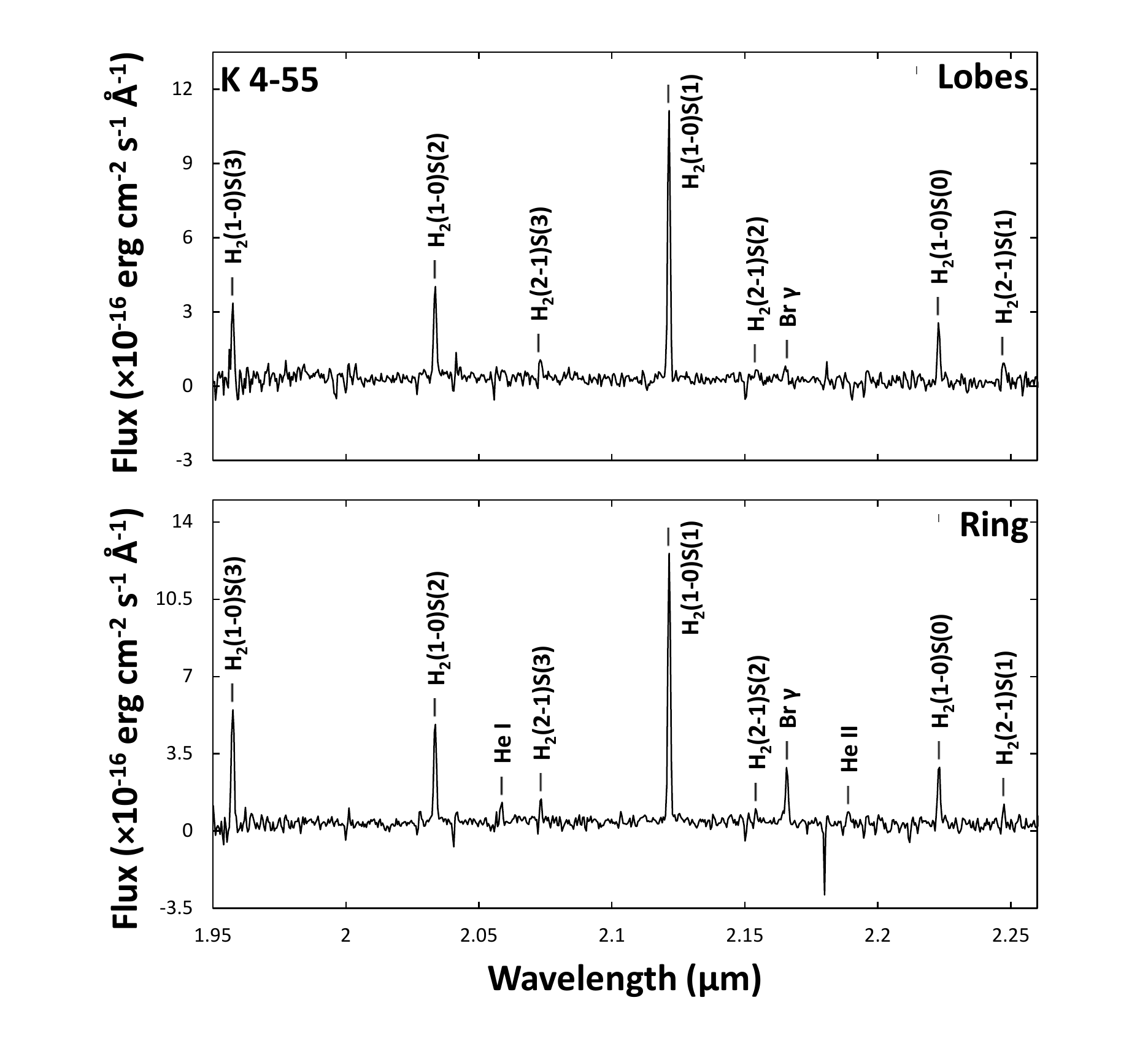}}
  \subfloat[M 1-75]{
  \includegraphics[trim = 10mm 36mm 10mm 15mm, clip, width=0.45\textwidth]{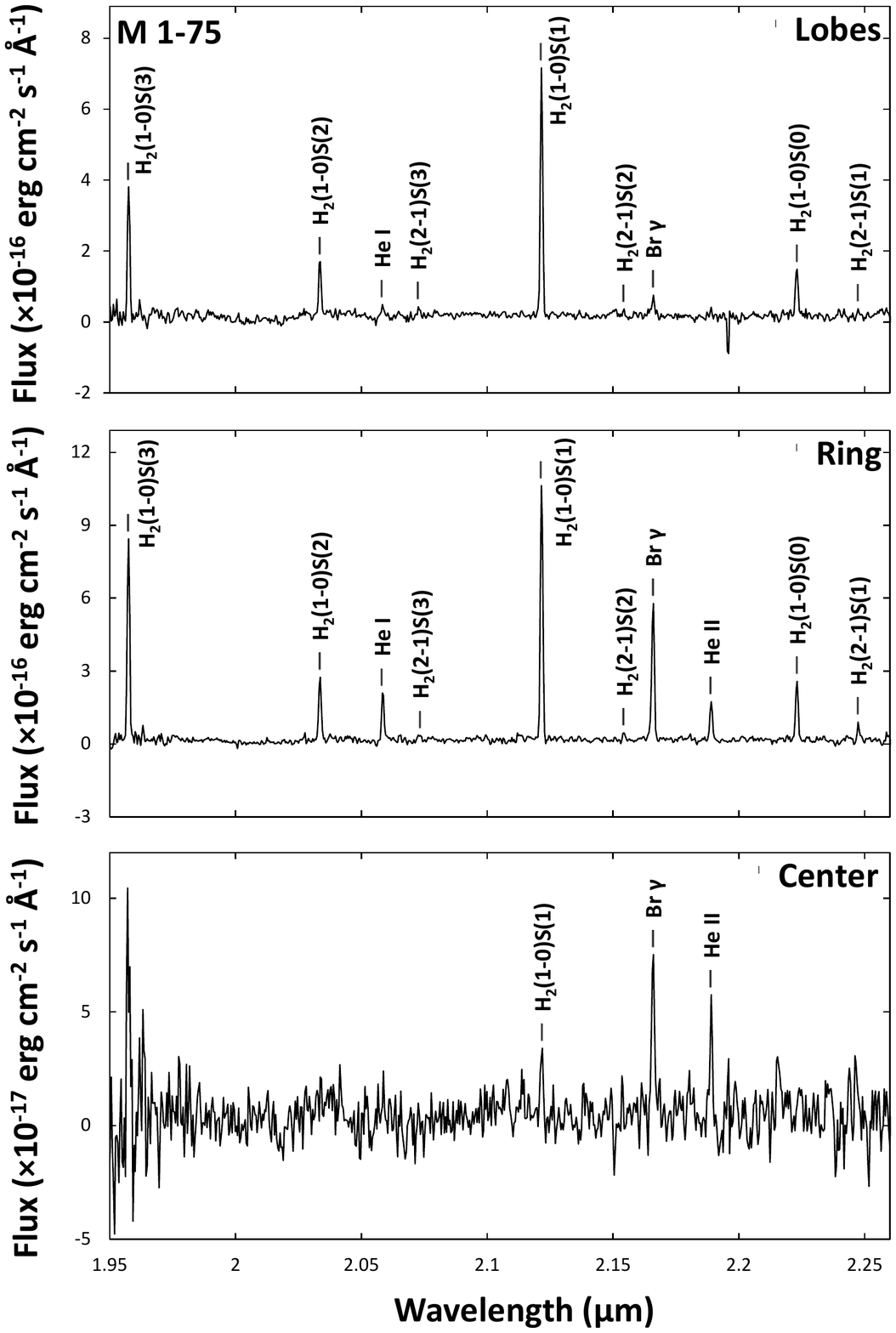}}
  \hspace{0.05\linewidth}
 \subfloat[M 1-92]{
  \includegraphics[trim = 10mm 20mm 10mm 20mm, clip, width=0.45\textwidth]{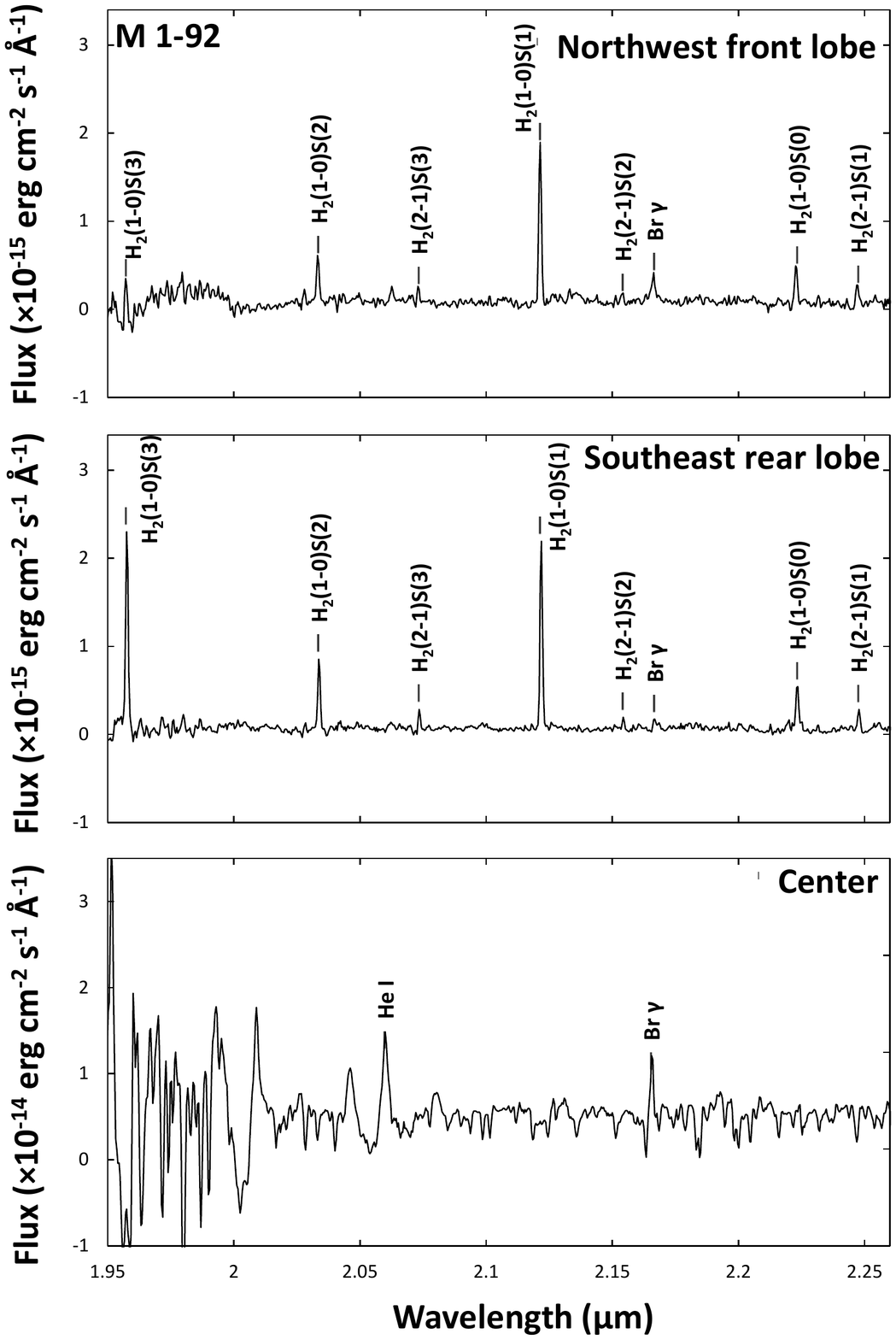}}
 \subfloat[M 4-17]{
  \includegraphics[trim = 10mm 20mm 10mm 20mm, clip, width=0.45\textwidth]{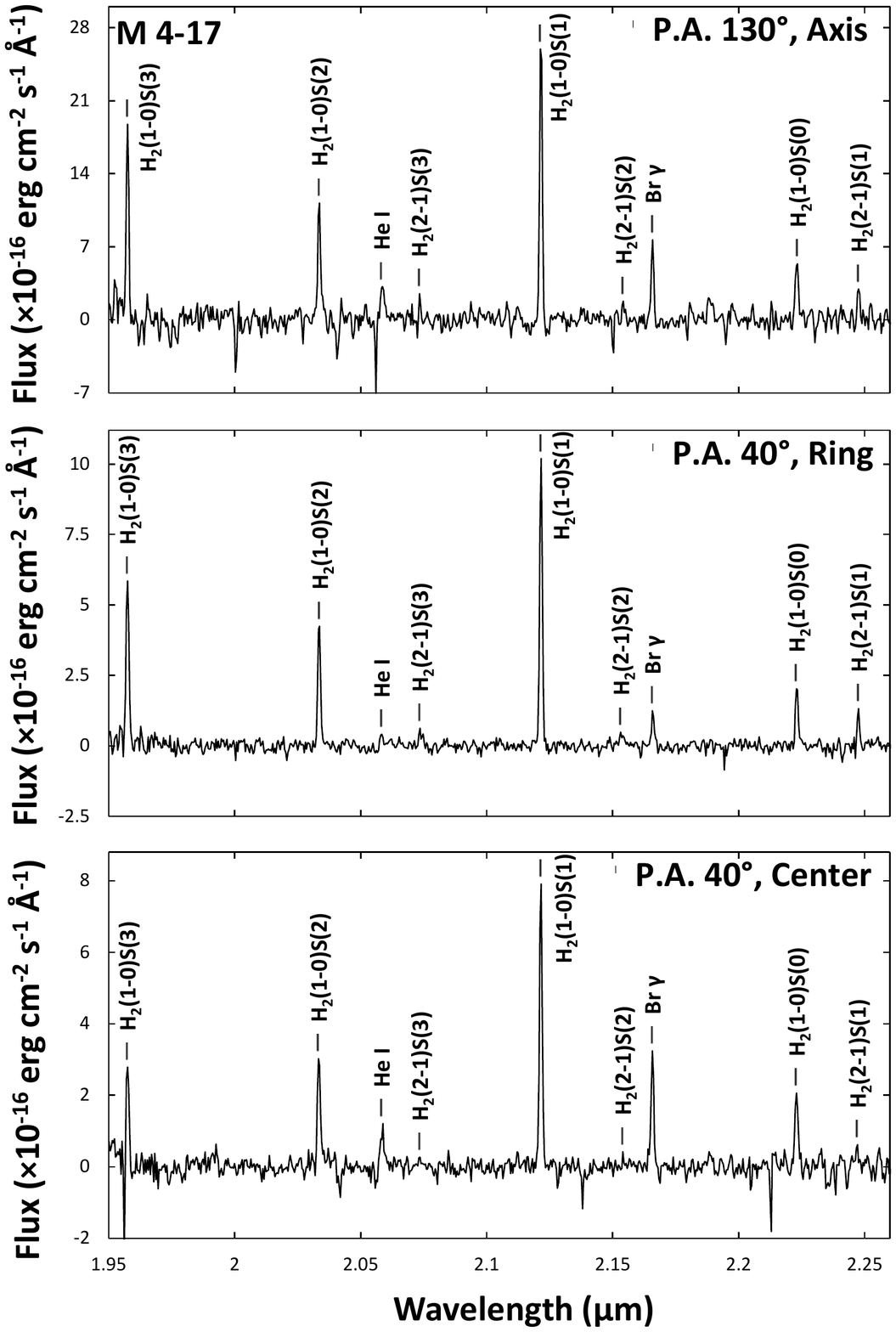}}
\caption{
TNG NICS $K$-band spectra of different regions of the H$_2$-bright bipolar 
PNe K\,4-55 (\emph{a}), M\,1-75 (\emph{b}), M\,1-92 (\emph{c}), and 
M\,4-17 (\emph{d}).  
The detected lines are labeled on the spectra.  
}
\label{spec_H2}
\end{figure*}


\subsection{Data analysis}

The spectra apertures were selected on the basis of the nebular 
near-IR morphologies (Figures~\ref{rendijas1} and  \ref{rendijas2}) and spatial profiles of 
selected spectral emission lines to probe different nebular regions 
and structures.  
Additionally, available optical images and kinematical information 
were used to select these apertures.

The PNe K\,4-55, M\,1-75, and M\,4-17 (Figure~\ref{rendijas1}) are relatively extended sources, 
with angular sizes many times that of the slit width.  
They lack a clear identification of their central star.  
As for K\,4-55, the slit PA was selected at 0$\degr$, covering the 
relatively bright central ring (R) and the pole-on bipolar lobes 
(L) at the position where they intersect \citep{1996ApJ...456..651G}.
M\,1-75 has a quadrupolar morphology \citep{Manchado96,2010A&A...519A..54S}, 
with two pairs of bipolar lobes aligned along PA 10$\degr$ and --30$\degr$, 
i.e.\ their main axis orientations differ by $\sim$40$\degr$. 
In this case, the slit was oriented at PA 60$\degr$, along a direction of 
bright emission in the central ring and relatively bright emission from 
one of the pairs of bipolar lobes outside the ring.
Three apertures were used to extract spectra corresponding to the 
central region (C), equatorial ring (R), and bipolar lobes (L).  
Finally, as for M\,4-17, the observations were acquired using two slits 
along perpendicular PAs 40$\degr$ and 130$\degr$.
The spectrum from the first slit, along the equatorial direction, allowed us 
to extract one-dimensional spectra of the central region (C) and equatorial 
ring (R).
The second slit goes along the nebular bipolar axis, where 
the emission is weaker than along the equatorial direction.  
In this case, all the emission detected in the slit was added into a 
single one-dimensional spectrum (A) which is representative of the 
whole nebula.

In contrast with the sources described above, Hb\,12, Hu\,2-1, IC\,4997,
and M\,1-92 (Figures~\ref{rendijas1} and  \ref{rendijas2}) have smaller angular sizes, of the order of the slit width.
Among these four sources, Hb\,12 shows the most detailed near-IR structure,
which can be described as a series of nested equatorial arcs surrounding a
dense core in an eye-shaped structure at the center of a more extended
hourglass nebula \citep{1996ApJ...461..288H, 1988ApJ...327L..27D,
1999ApJ...522L..69W, 2000ASPC..199..267H, 2014AJ....148...98C}.
The slit was oriented along the bipolar axis at PA=$-$5$\degr$, covering the
core (C), core envelope (E), and ring (R). The optical structure of
Hu\,2-1 is
complex, with a central equatorial ring containing a small elliptical
shell, a
pair of bipolar lobes, two pairs of collimated bipolar knots, and outer
structures \citep{2001MNRAS.321..487M}. Radio continuum observations of
IC\,4997 reveal an outer bipolar shell and an inner elliptical one
\citep{1998ApJ...496..274M}. Otherwise, both Hu\,2-1 and IC\,4997 are rather
compact in the near-IR images shown in Figure~\ref{rendijas2}, with relatively
bright emission at their central regions and much fainter extended emission.
The slit PAs were selected to cover the main nebular axis in the case of
Hu\,2-1, and the minor nebular axis in the case of IC\,4997, but in view of
their inconspicuous near-IR morphology, one-dimensional spectra have been
extracted for the central regions (C) and extended emission (E).
Finally, M\,1-92, the Minkowski's Footprint Nebula, has a clear
bipolar morphology, with the nebular axis oriented along PA 313$\degr$
\citep{1996ApJ...468L.107T,1998A&A...331..361B}.
The Southeast lobe is fainter than the Northwest one because the
inclination of
the nebula causes differential extinction by an extended equatorial disk.
The slit was placed along the nebular axis, covering the Northwest (front)
lobe (FL), the central region (C), and the Southeast (rear) lobe (RL).

\section[]{One-dimensional spectra}

\begin{figure*}
  \centering
  \subfloat[Hb 12]{
  \includegraphics[trim = 20mm 36mm 20mm 10mm, clip, width=0.45\textwidth]{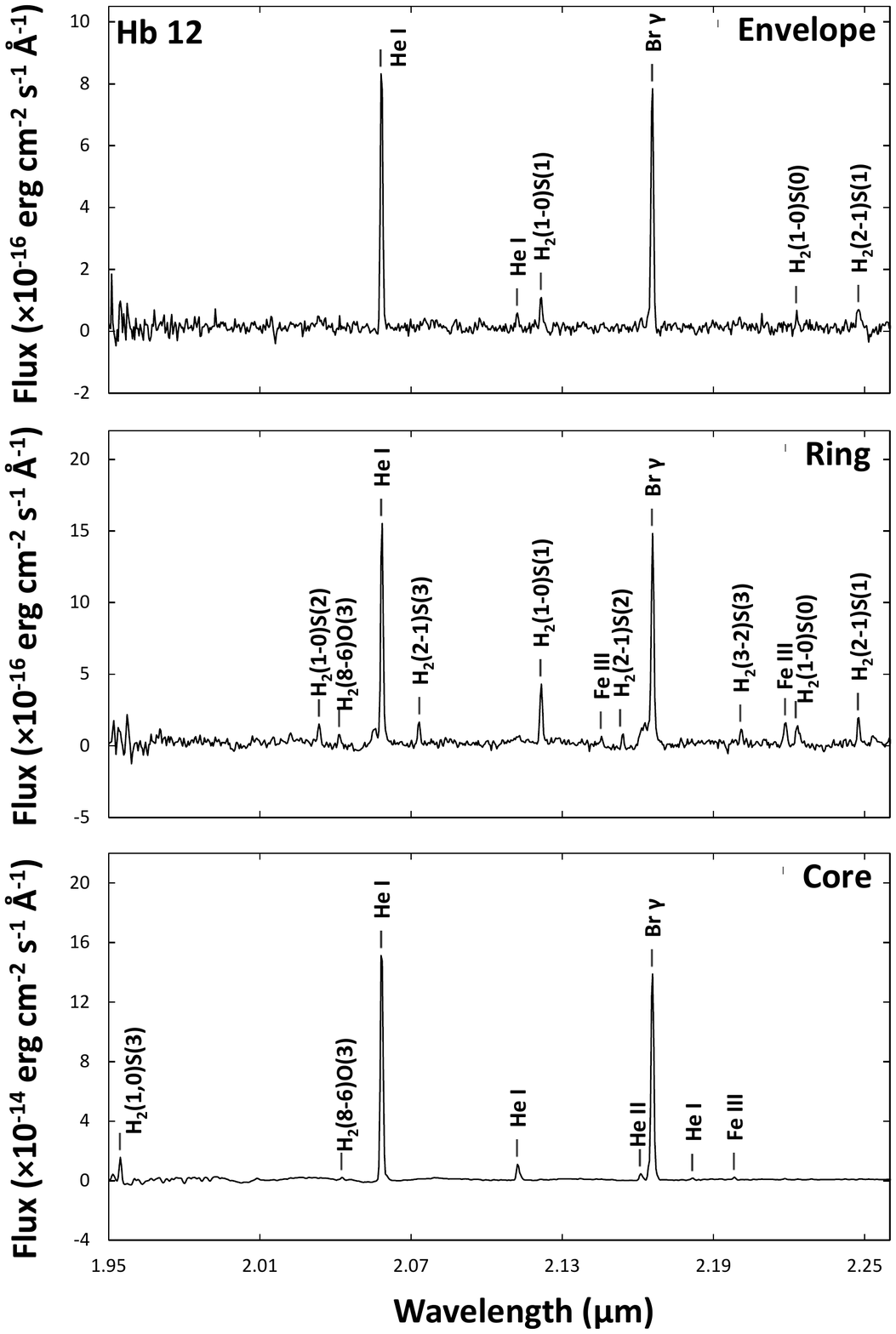}}
  \hspace{0.01\linewidth}
  \subfloat[IC 4997]{
  \includegraphics[trim = 20mm 110mm 20mm 15mm, clip, width=0.45\textwidth]{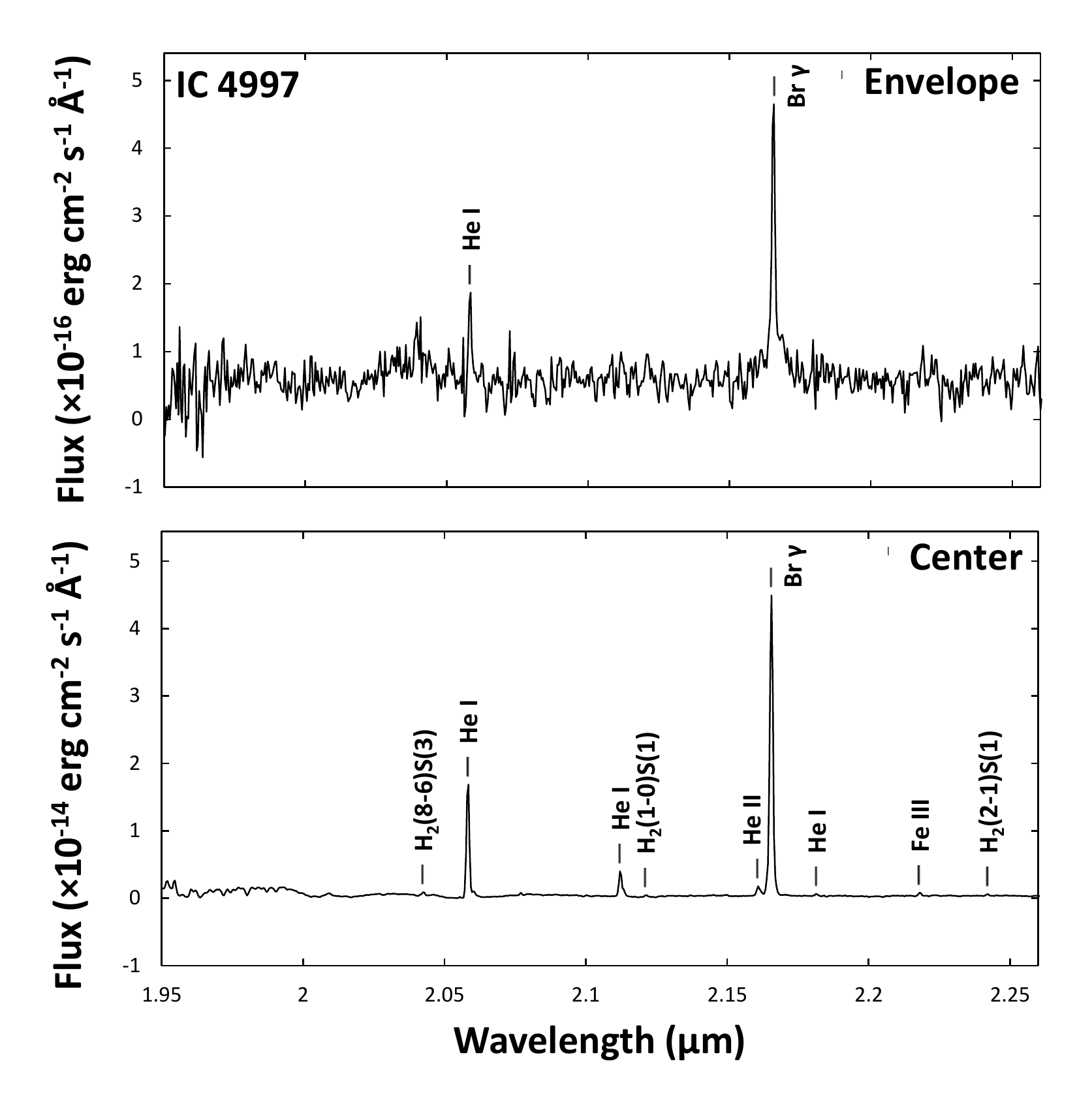}}
  \subfloat[Hu 2-1]{
  \includegraphics[trim = 20mm 110mm 20mm 15mm, clip, width=0.45\textwidth]{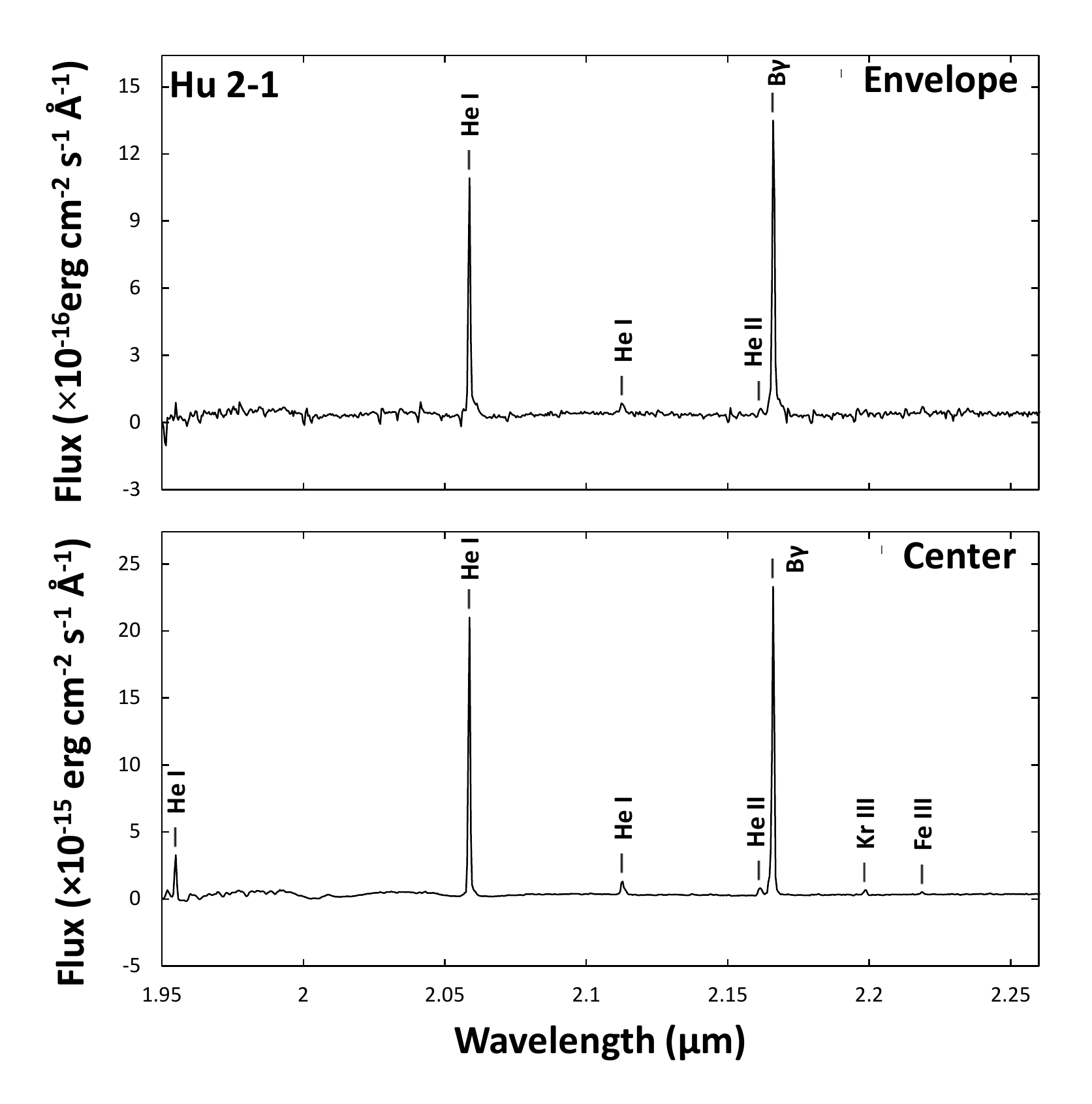}}
  \caption{\label{spec_ion}
TNG NICS $K$-band spectra of different regions of the H$_2$-weak PNe 
Hb\,12 (\emph{a}), IC\,4997 (\emph{b}), and Hu\,2-1 (\emph{c}).  
The detected lines are labeled on the spectra.  
}
\end{figure*}

The one-dimensional spectra of the selected regions in our sample 
of PNe are shown in Figures~\ref{spec_H2} and \ref{spec_ion}.  
The distribution of the different spectra in these two figures is not 
accidental: 
PNe in Figure~\ref{spec_H2} show spectra rich in emission lines of 
molecular hydrogen, whereas the spectra of PNe in Figure~\ref{spec_ion} 
have weak or absent H$_2$ lines, but bright H~{\sc i} and He~{\sc i} lines.  
Accordingly, we have split the PNe in our sample into two groups, 
the H$_2$-bright PNe K\,4-55, M\,1-75, M\,1-92, and M\,4-17, and the 
H$_2$-weak PNe Hb\,12, Hu\,2-1, and IC\,4997.

The line identifications and their rest wavelengths, and the normalized 
line intensities in the selected regions (see Figs.~\ref{spec_H2} and 
\ref{spec_ion}) of the H$_2$-bright PNe are listed in Table~\ref{spec_H2}, 
and those of the H$_2$-weak PNe in Table~\ref{spec_ion}.  

Line ratios have not been undereddened since the visual absorption 
towards these PNe imply $A_K<$0.15 mag, which result in line relative 
intensities corrections in the full $K$-band smaller than 5\%.  
The results in these tables confirm the segregation of the PNe in our 
sample into two different groups.  
The H$_2$-bright PNe in Table \ref{spec_H2} have a large number of 
H$_2$ emission lines in the near-IR $K$-band and show H$_2$ 1-0 S(1) 
to Br$\gamma$ line ratios much larger than unity.  
On the contrary, the H$_2$-weak PNe in Table \ref{spec_ion} show 
a small number of H$_2$ emission lines, these are generally weak, 
and the Br$\gamma$ line is much brighter than the H$_2$ 1-0 S(1) 
line when the latter is detected.

The spectral properties of the H$_2$-bright and H$_2$-weak PNe are described 
in further detail in the next sections.

\subsection{H$_2$-bright BPNe}

The near-IR images of the H$_2$-bright BPNe K\,4-55, M\,1-75, and M\,4-17 
display large, well resolved, and bright ring-like structures and extended 
bipolar lobes.  
These are typical R-BPNe.  
The one-dimensional spectra of these PNe have been extracted from regions 
corresponding to the outer bipolar lobes, the central ring, and the interior 
of this ring.  
As for M\,4-17, the emission from the bipolar lobes was found too weak, 
and it has been merged into one single spectrum for the whole nebula.

On the other hand, M\,1-92 has relatively bright bipolar lobes, but the 
central region is not resolved into a ring, but it is better described 
as a bright core-like feature, i.e., it is a W-BPN.  
In this case, the one-dimensional spectra correspond to the bipolar lobes 
and the core.  
The spectra of the two bipolar lobes have been considered separately 
as they show differing surface brightness and line ratios.

The bipolar lobes of these sources are characterized by extreme ratios 
between the H$_2$ 1-0 S(1) and Br$\gamma$ emission lines, with values 
ranging from $\sim$7 (rear lobe of M\,1-92) up to $\sim$14 (K\,4-55). 
Other lines from ionized species (He~{\sc i} and He~{\sc ii}) are 
equally faint.  
Otherwise, several emission lines from other H$_2$ transitions are quite 
bright.

The ring-like structures of K\,4-55, M\,1-75, and M\,4-17 still display 
numerous and bright H$_2$ emission lines, but their intensity, relative 
to that of Br$\gamma$ and He~{\sc i} and He~{\sc ii} lines, is lower.  
As for the innermost region of M\,1-92, the H$_2$ lines are undetected.  
This is also the case for the central regions of M\,1-75, whereas, in 
the case of M\,4-17, H$_2$ lines are still present, but weak.  
The projection of the bipolar lobes on the innermost regions of M\,4-17 
would explain the detection of H$_2$ emission there.

The above description of the near-IR $K$-band spectra of H$_2$-bright 
BPNe suggests a scenario where the innermost regions of these nebulae 
are dominated by ionized material, whereas molecular material is found 
(and excited) in ring-like structures and bipolar lobes.  
Material close to the central star 
exhibits higher ionization degrees, as implied by the intensity 
of the He~{\sc i} and even He~{\sc ii} emission lines, 
but molecular material can still shield from the stellar UV radiation 
in clumps at ring- or torus-like structures or survive in the bipolar 
lobes where the UV radiation flux has been diluted by the large 
distance towards the central star.

\subsection{H$_2$-weak BPNe}

The PNe in this subsample, Hb\,12, Hu\,2-1, and IC\,4997, are all 
dominated in the near-IR by the bright emission from an unresolved 
core (Fig.~\ref{obs_spec}).  
The outermost emission in these nebulae is much weaker: 
the eye-shaped feature around the bright core of Hb\,12, 
which has been described as the line where the outer 
bipolar lobes join \citep{2000ASPC..199..267H}, 
the weak bipolar lobes of Hu\,2-1, and 
the outer emission from a knot-like feature in IC\,4997.  
Hb\,12 can be classified as W-BPN, whereas Hu\,2-1 is better described 
as a R-BPN.  
The detailed classification of IC\,4997 is uncertain.  

\begin{figure*}
  \centering
  \includegraphics[trim = 20mm 20mm 20mm 20mm, clip, width=0.5\textwidth]{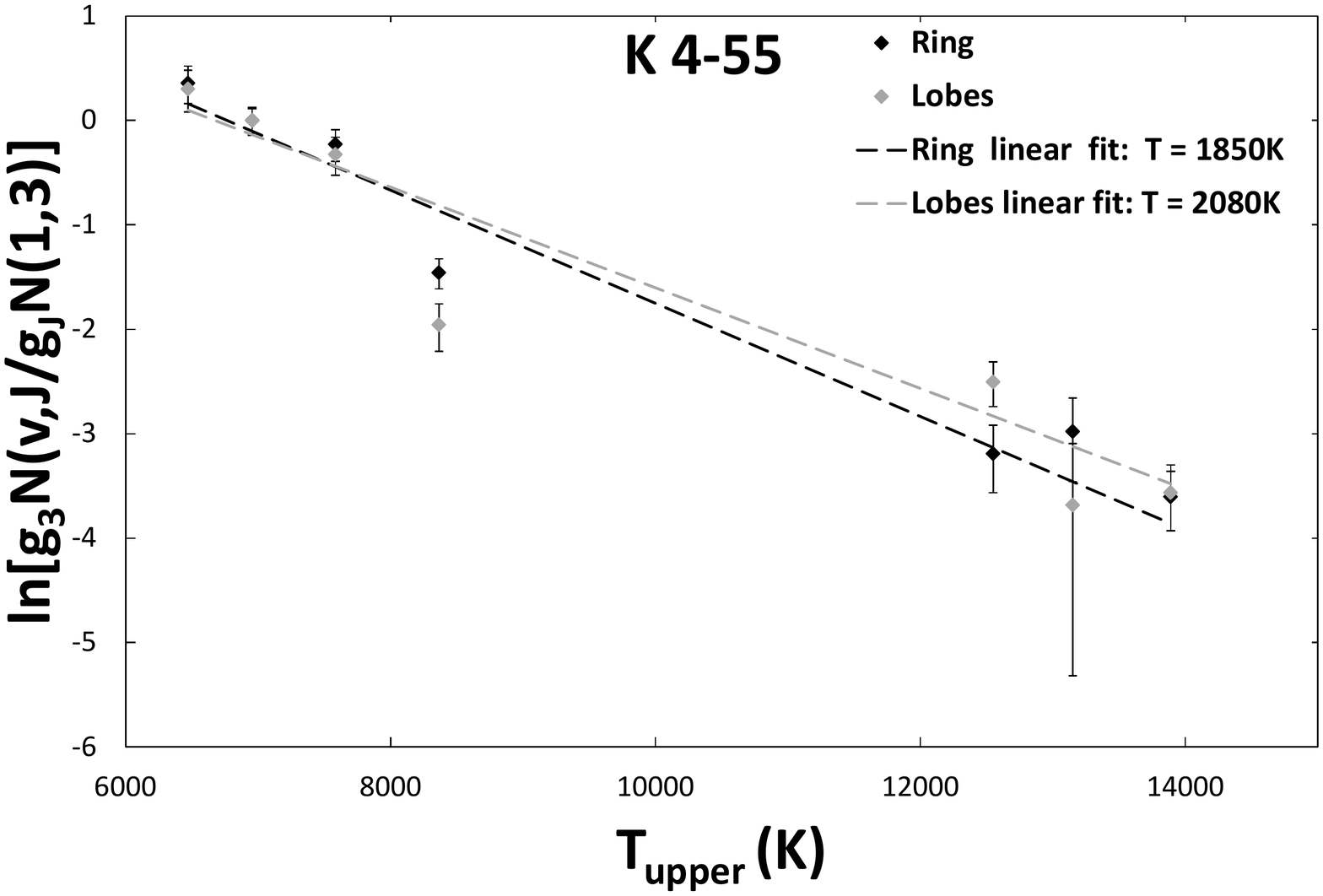}
  \includegraphics[trim = 20mm 20mm 20mm 20mm, clip, width=0.5\textwidth]{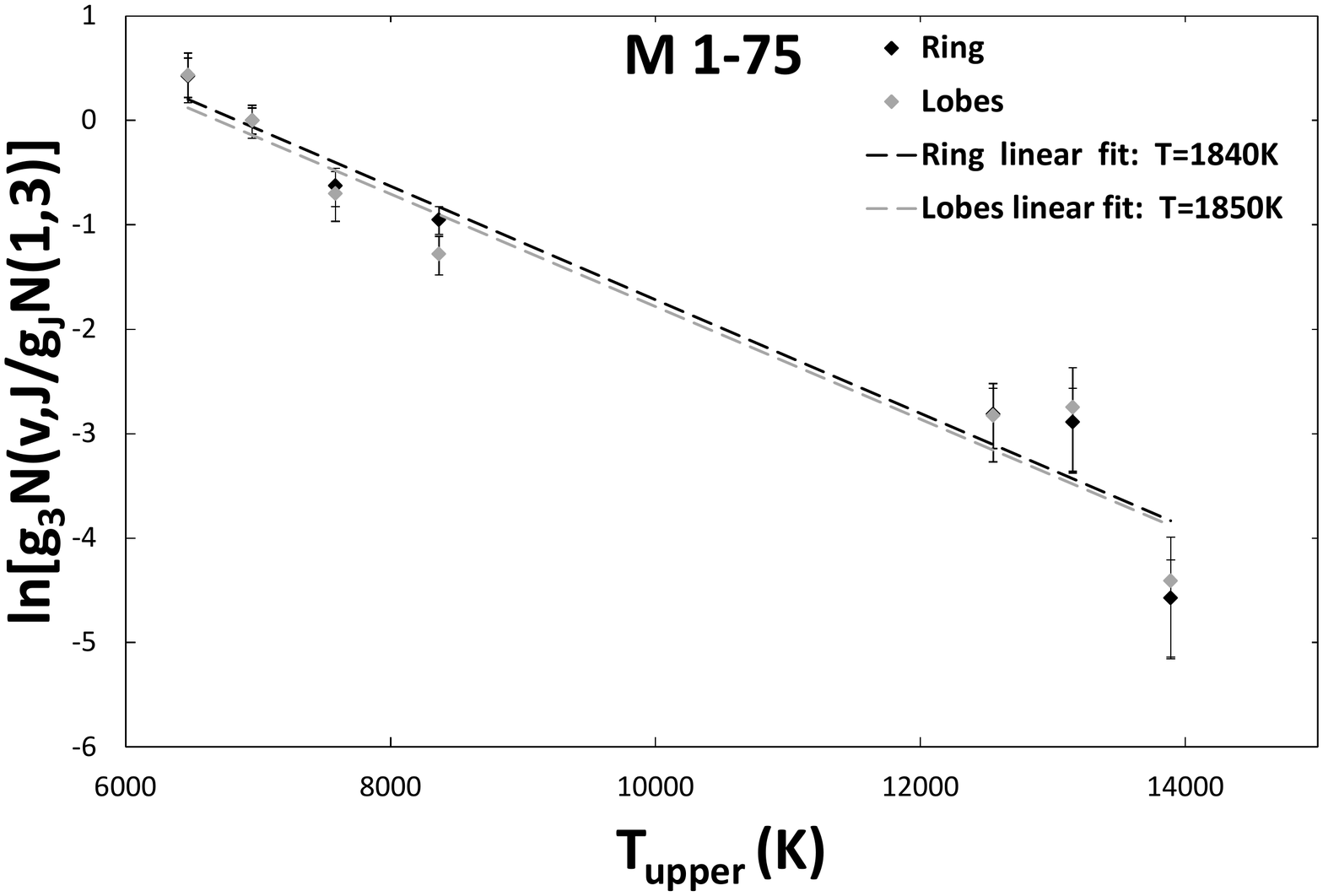}
  \hspace{0.01\linewidth}
  \includegraphics[trim = 20mm 20mm 20mm 22mm, clip, width=0.5\textwidth]{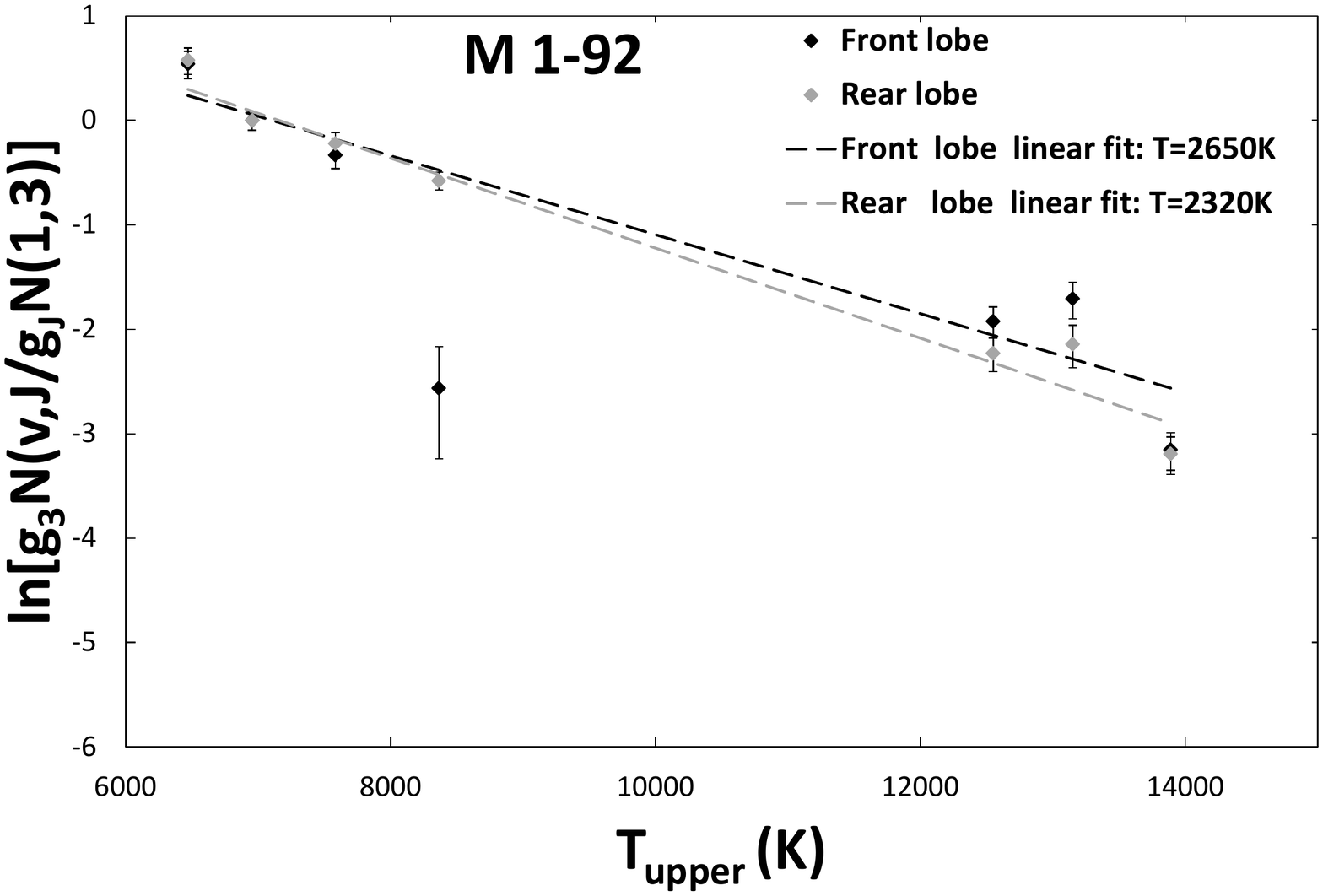}
  \includegraphics[trim = 20mm 20mm 20mm 22mm, clip, width=0.5\textwidth]{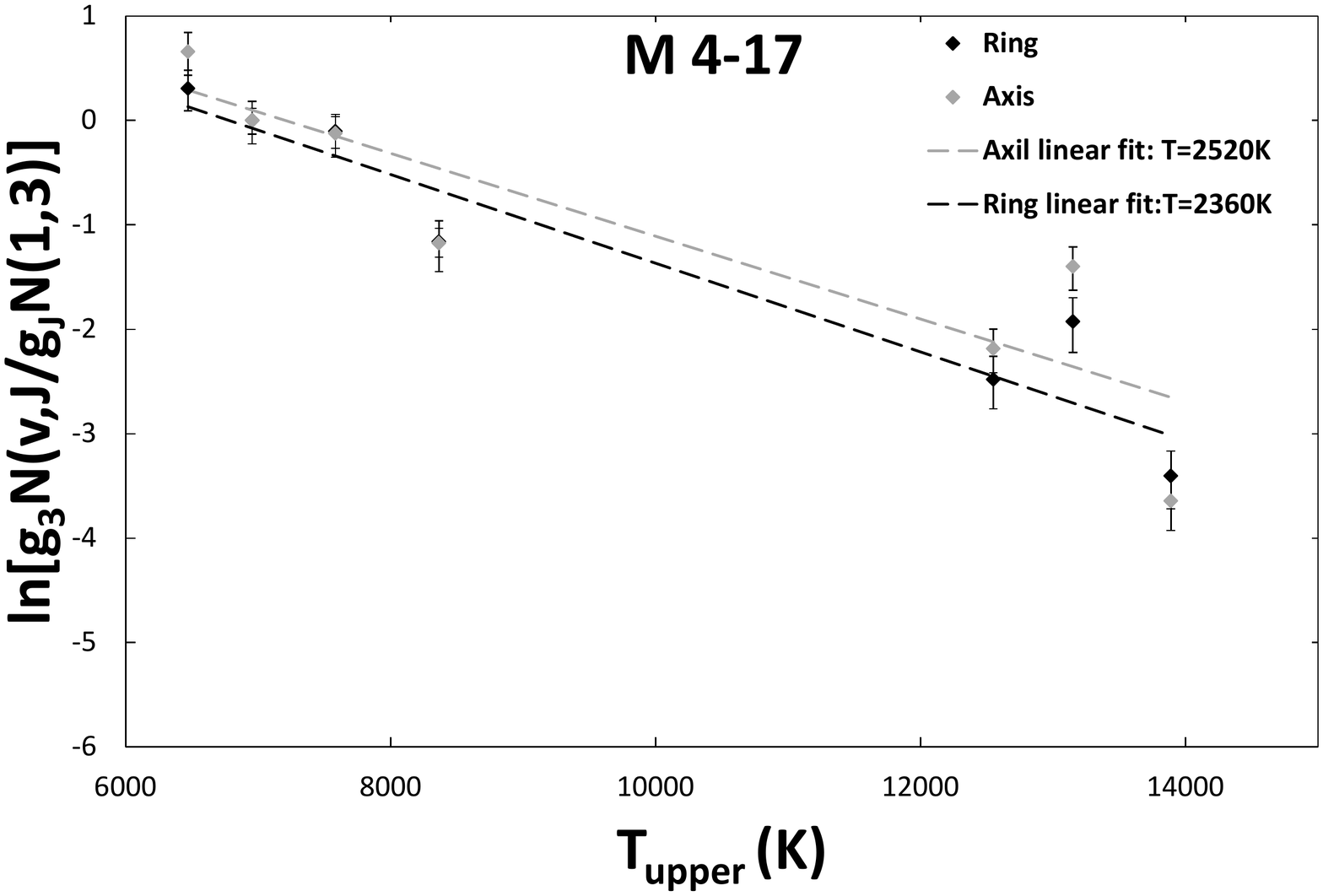}
  \caption{
Excitation diagrams for the H$_2$ lines in the bipolar PNe K\,4-55 
(top-left), M\,1-75 (top-right), M\,1-92 (bottom-left), and M\,4-17 
(bottom-right).  
Data derived for different regions are plotted with different grey 
intensities.  
The linear fits to the different data sets, implying the temperature 
for thermal excitation, are labeled. 
}
\label{temperature}
\end{figure*}

The near-IR $K$-band spectra from the central regions of these sources 
are dominated by the Br$\gamma$ and He~{\sc i} $\lambda$2.0580 $\mu$m 
lines.  
Other He~{\sc i} and even the He~{\sc ii} $\lambda$2.1614 $\mu$m 
lines are also clearly detected in the spectra of these compact 
regions.  
Weak Fe~{\sc iii} $\lambda$2.1460 $\mu$m, Fe~{\sc iii} $\lambda$2.2184 
$\mu$m, and Kr~{\sc iii} $\lambda$2.1993 $\mu$m lines are detected in 
these PNe.  
The H$_2$ lines, if present in these spectra, are not that prominent 
as in the H$_2$-bright PNe.  

The spectra of the outer regions of these nebulae have much lower 
signal-to-noise ratio than those of their inner regions.  
Hu\,2-1 shows Br$\gamma$ and He~{\sc i} and He~{\sc ii} lines, 
whereas the spectrum of the outer emission in IC\,4997 only 
displays the Br$\gamma$ and He~{\sc i} $\lambda$2.0580 $\mu$m 
line.  
The eye-shaped feature of Hb\,12 shows the most interesting spectrum, 
with a number of H$_2$ and Fe~{\sc iii} lines in addition to the 
Br$\gamma$, He~{\sc i} $\lambda$2.0580 $\mu$m, and He~{\sc ii} 
$\lambda$2.1614 $\mu$m lines. Similar to the results reported by 
\citet{1996ApJ...461..288H} and \citet{1999ApJS..124..195H}. 

To conclude, Hu\,2-1 and IC\,4997 show very little evidence for the 
presence of molecular material, whereas Hb\,12, although dominated 
by emission from ionized material, shows some evidences of molecular 
material in the outer regions.

\section{H$_2$ excitation}

The near-IR $K$-band includes a number of emission lines from 
ro-vibrational transitions of the H$_2$ molecule.  
In particular, the emission lines of several transitions from the 
vibrational levels 2--1 and 1--0 are very prominent in this band.  
The H$_2$ 1--0~S(1) to 2--1~S(1) and 3--2~S(3) line ratios have been 
traditionally used to diagnose the excitation mechanism of the H$_2$ 
molecules.  
If shock excitation is dominant, then the 1--0~S(1)/2--1~S(1) and 
1--0~S(1)/3--2~S(3) line ratios are expected to be high, 10--20 
and 10--100, respectively, 
although lower values of the 1--0~S(1)/2--1~S(1) line ratio, down 
to $\sim$4, have also been proposed \citep{SH78,1995A&A...296..789S}.
Even lower values of the 1--0~S(1)/2--1~S(1) line ratio, down to $\sim$4, 
can be expected in shocked molecular clouds \citep{SH78}, where physical 
conditions may be not similar to those found in PNe.  
For the case of UV excitation, as in photo-dissociation regions, 
models predict 1--0~S(1)/2--1~S(1) and 1--0~S(1)/3--2~S(3) 
line ratios of 2--3 and $\sim$8, respectively 
\citep{1990ApJ...365..620B,1995ApJ...455..133H}.  
It must be noted, however, that for high densities, above 10$^5$ cm$^{-3}$, 
collisional de-excitation can thermalize the low-energy states of UV 
excited H$_2$ molecules, thus producing ratios for these lines close to 
those expected in cases of pure shock excitation \citep{1995ApJ...455..133H}.

In our sample, we can recognize two different groups among those that 
show H$_2$ emission lines.  
The H$_2$-bright PNe in Table \ref{spec_H2} exhibit large H$_2$ 
1--0~S(1)/2--1~S(1) line ratios, from $\sim$5.0 in M\,1-92 to 
$\sim$18 in the ring of K\,4-55.  
The H$_2$ 3--2~S(3) line at $\lambda$2.2104 $\mu$m is not detected, 
implying large H$_2$ 1--0~S(1)/3--2~S(3) line ratios.  
These large line ratios are indicative of shock excitation 
as collisional de-excitation can be discarded because the electron 
densities of these sources are much lower than those required 
($>$10$^5$ cm$^{-3}$) for this mechanism to be efficient: 
500 cm$^{-3}$ for K\,4-55 \citep{1996ApJ...456..651G}, 
1100 cm$^{-3}$ for M\,1-75 \citep{GSM95}, 
2300 cm$^{-3}$ for M\,1-92 \citep{Solf94}, 
100--1000 cm$^{-3}$ for M\,4-17 \citep{Kaler96}.
On the other hand, the same line ratios implied by the data in 
Table~\ref{spec_ion} for the H$_2$-weak PNe are much smaller, 
with H$_2$ 1--0~S(1)/2--1~S(1) in the range 0.8--2.1 for IC\,4997 and 
Hb\,12.  
The H$_2$ 3--2~S(3) line is only detected in the region enveloping the 
core of Hb\,12, for a 1--0~S(1)/3--2~S(3) line ratio $\sim$3.3.  
We thus conclude that in all cases of H$_2$-weak PNe where the H$_2$ 
lines are detected, there is clear evidence of UV excitation.

The population of the H$_2$ molecule energy levels provides 
an accurate assessment of the excitation conditions of these 
molecules \citep[e.g.][]{1994ApJ...437..281H,1999ApJS..124..195H}. 
In the case of pure shock or collisional excitation, a single-temperature 
Boltzmann distribution describes the population of the energy levels of 
the H$_2$ molecule.  
Then, the value of the vibrational excitation temperature, $T_{ex}$($\nu$), 
is coincident with the rotational excitation temperature, $T_{ex}$(J).  
In the case of UV excitation, the population of the energy levels 
departs from a Boltzmann distribution and the vibrational and 
rotational excitation temperatures differ, $T_{ex}$($\nu$)$\neq T_{ex}$(J).

To assess whether excitation is purely collisional or due to UV fluorescence, 
the H$_2$ line ratios from the same ro-vibrational ground levels have been 
used to derive the column densities relative to the $\nu$= 1 and J=3 state 
plotted in Figure~\ref{temperature}.  
The data points in these plots can be fitted by a single line, implying 
that the population of the energy levels follows a Boltzmann distribution 
and thus that excitation is mostly collisional.   
A reduced $\chi^2$ fit to these data points indeed confirms that a 
single line adequately decribes their distribution, with reduced 
$\chi^2$ values between 0.8 and 1.8. 
The excitation temperature can be inferred from the slope of the line 
fitted to the data points: 
$T_{ex}$=1850--2080 K for K\,4-55, 
$T_{ex}$=1840--1580 K for M\,1-75, 
$T_{ex}$=2320--2650 K for M\,1-92, and 
$T_{ex}$=2360--2520 K for M\,4-17.

\begin{figure}
  \centering
  \includegraphics[trim = 20mm 20mm 0mm 20mm, clip, width=0.5\textwidth]{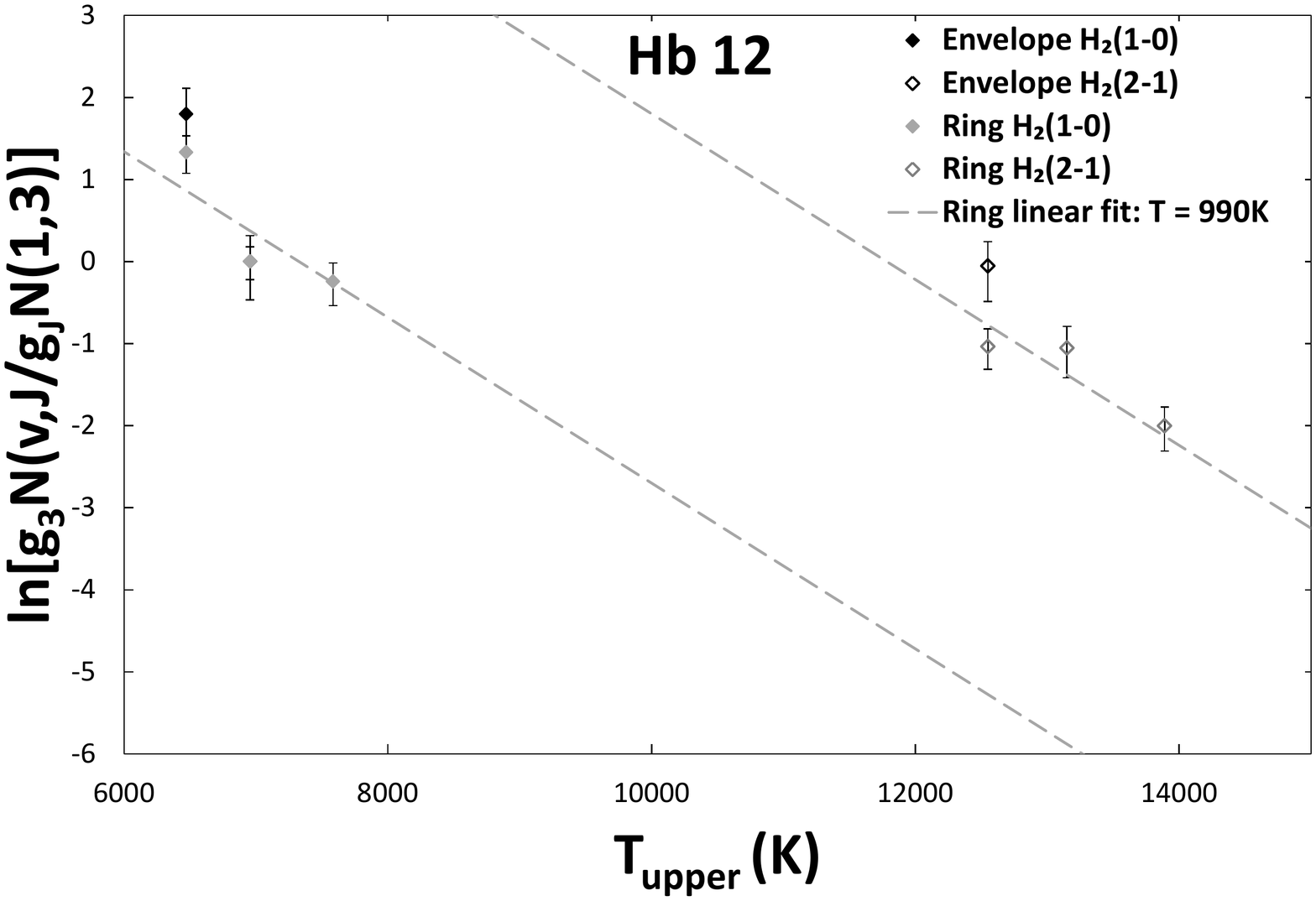}
  \caption{
Excitation diagrams for the H$_2$ lines in the bipolar PNe Hb\,12. 
Data derived for different regions are plotted with different symbols and grey 
intensities.  
}
\label{temp_hb12}
\end{figure}

As for Hb\,12, the H$_2$ line ratios are consistent with those 
presented by \citet{1996ApJ...461..288H} and \citet{LR96} who 
concluded that the excitation mechanism of the H$_2$ molecules 
in Hb\,12 is pure fluorescent.  
The available line ratios have been used to derive the column densities 
relative to the $\nu$= 1 and J=3 state plotted in Figure~\ref{temp_hb12}.  
In this case, 
the reduced $\chi^2$ for a single linear fit is $\simeq$10, which indicates 
that the population distribution is not thermal.  
Different linear fits describe much more accurately the data 
points from different vibrational levels (reduced $\chi^2\simeq$1.3), 
thus implying fluorescence excitation.  
The excitation temperature, $\simeq$990 K, is consistent with that of 
1395$\pm$450 K reported by \citet{1996ApJ...461..288H}.  

Finally, for IC\,4997, insufficient emission lines were detected for an analysis 
of temperature, whereas no H$_2$ is detected for Hu\,1-2.

\section{Discussion}


A significant fraction of PNe display H$_2$ emission lines.  
Most of them have bipolar morphologies, leading \citet{1994ApJ...421..600K} 
to state the so-called Gatley's rule, for which the detection of the 2.1218 
$\mu$m S(1) H$_2$ line was considered to be conclusive proof of the bipolar 
nature of a PN.  
As more sensitive observations are been carried out, many exceptions 
to this rule have been reported. 
Large, very evolved PNe of different morphological groups have been 
found to display emission in H$_2$ lines \citep{2013MNRAS.429..973M}.  
There is not a tight correlation between occurrence of H$_2$ 
emission and bipolar morphology in fluorescence-excited sources 
\citep{2002A&A...387..955G}.  
The general consensus nowadays is that emission of H$_2$ lines is not 
exclusive of PNe with bipolar morphology, although they tend to be 
brighter in H$_2$ than other PNe \citep{2013MNRAS.429..973M}.

In this work we have presented near-IR $K$-band intermediate-resolution 
spectroscopic observations of a sample of bipolar and/or elongated PNe.  
These observations detect H$_2$ emission lines in Hb\,12, K\,4-55, M\,1-75, 
M\,1-92, and M\,4-17.  
Very weak H$_2$ emission is also detected in IC\,4997. 
Despite its bipolar morphology, no H$_2$ emission is detected in Hu\,2-1.  
The high detection rate is not surprising, giving the selection criteria, and 
it certainly emphazises the higher H$_2$ brightness of bipolar PNe.

\begin{figure*}
  \centering
  \includegraphics[trim = 20mm 20mm 90mm 0mm, clip, width=0.5\textwidth]{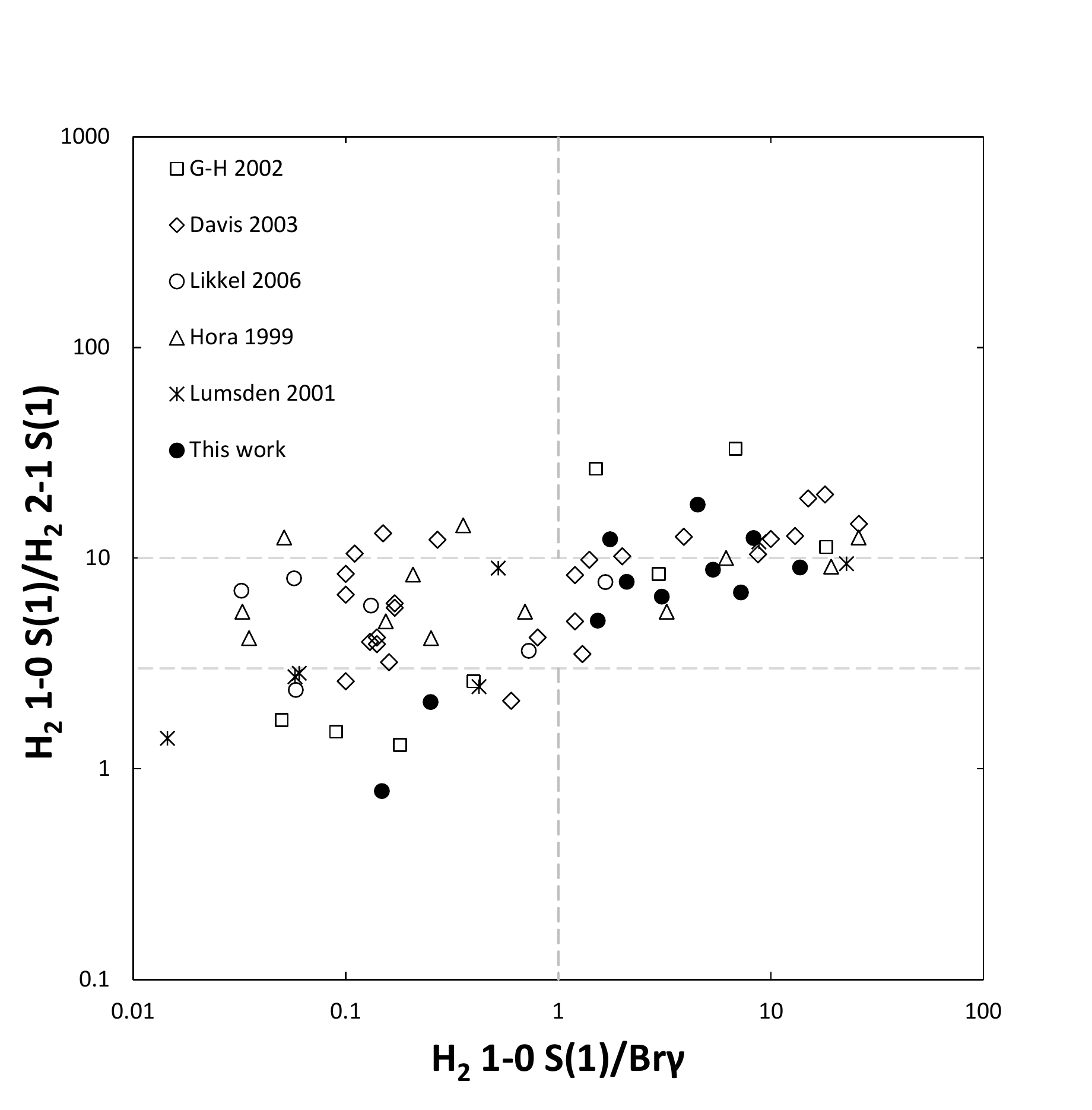}
  \includegraphics[trim = 20mm 20mm 90mm 0mm, clip, width=0.5\textwidth]{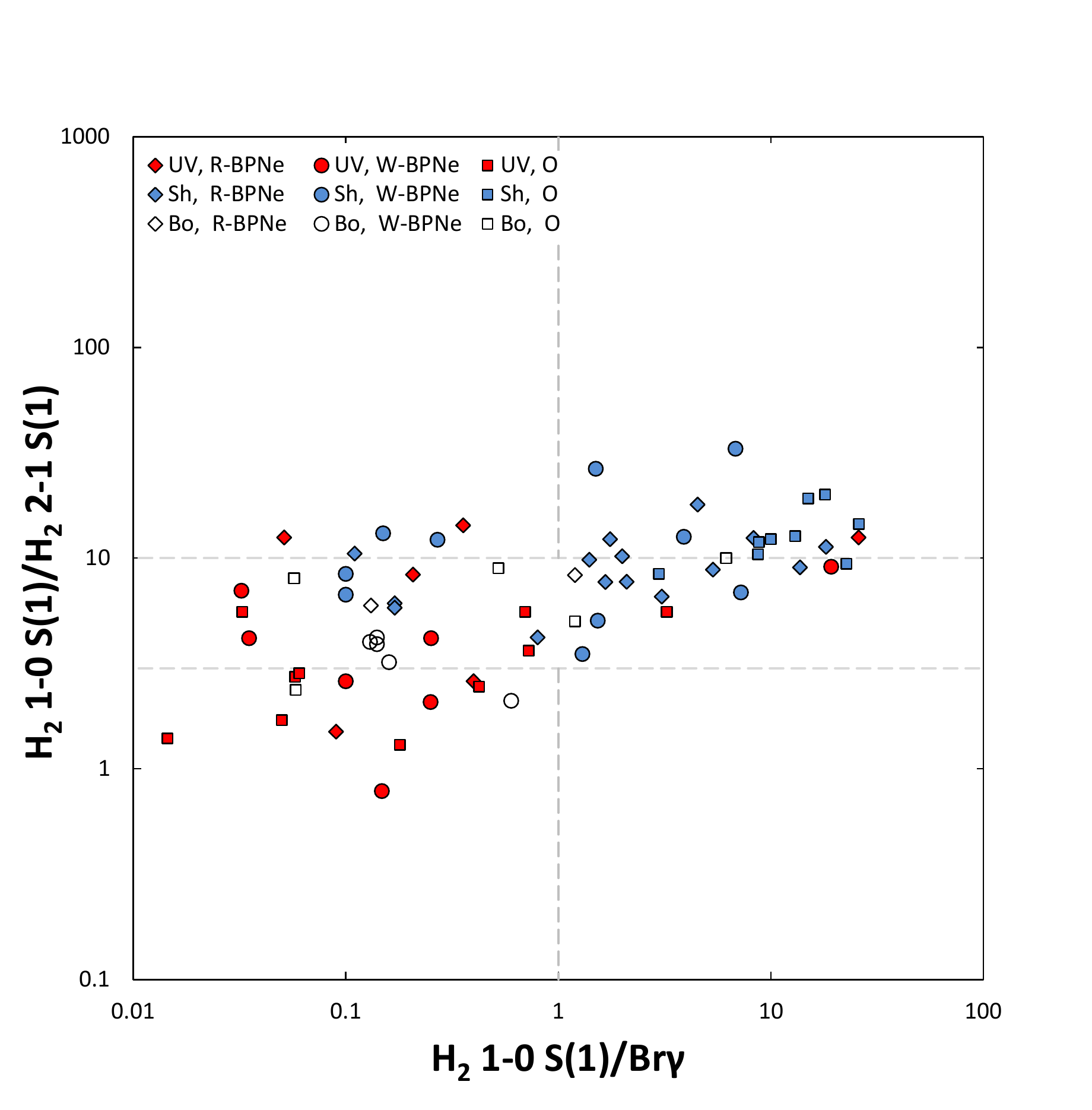}
  \caption{
Distribution of PNe in the H$_2$ 1-0 S(1)/H$_2$ 2-1 S(1) versus 
H$_2$ 1-0 S(1)/Br$\gamma$ line ratios diagram.  
{\it (left)} The sample of PNe with available measurements of these line ratios 
according to the original reference.  
{\it (right)} Same as above, but marking the excitation mechanism and detailed 
bipolar morphology as waist (W-BPNe) or ring (R-BPNe).  
Sources with unclear, unknown, or other morphological types are also 
marked.  
Sources reported to have a mixture of shock and UV excitation are also 
singled out.  
For comparison, the horizontal lines mark the limits for UV-excited ($<$3) 
and shock-excited ($>$10) H$_2$ typically assumed in the literature.  
The vertical line at H$_2$ 1-0 S(1)/Br$\gamma$ unity separates 
Br$\gamma$-bright sources (to the left) from H$_2$-bright sources
(to the right).  
}
\label{plot_H2Brg}
\end{figure*}

Among our sample, there is a clear correlation between detailed bipolar 
morphology and relative strength of the H$_2$ emission: 
R-BPNe, with well developed equatorial rings and large bipolar lobes 
tend to exhibit larger H$_2$ to Br$\gamma$ line ratios than W-BPNe 
with narrow waists at their equatorial regions, 
with the possible exception of Hu\,2-1, although we note that its 
equatorial ring has not developed as those of M\,1-75 and M\,4-17.  
This may suggest not only that equatorial rings offer a shelter for 
H$_2$ molecules, but that shock excitation, which is a much more 
efficient excitation mechanism than UV fluorescence and thus produces 
higher levels of emission in the H$_2$ lines, is the dominant excitation 
mechanism in these objects.  
Indeed, our spectroscopic observations confirm that shock-excitation is 
prevalent among R-BPNe.  

This trend is investigated into more detail in Figure~\ref{plot_H2Brg}, 
where the H$_2$ 1-0 S(1)/H$_2$ 2-1 S(1) and H$_2$ 1-0 S(1)/Br$\gamma$ 
line ratios are plotted for a sample of objects available in the literature
\citep[this paper;][]{1999ApJS..124..195H,2001MNRAS.328..419L,2002A&A...387..955G,2003MNRAS.344..262D,2006AJ....131.1515L}.  
The distribution of W-BPNe or R-BPNe in this plot confirms the previous 
trend: R-BPNe outnumber W-BPNe by two to one 
(12 R-BPNe for 7 W-BPNe) for sources brighter in H$_2$ than in Br$\gamma$, 
but this ratio reverses for Br$\gamma$-brighter sources 
(6 R-BPNe for 13 W-BPNe for sources with H$_2$/Br$\gamma<$0.5).  
We also see here trends previously reported on the effects of nebular 
evolution on the excitation mechanism (and thus in the level) of H$_2$ 
emission.  
In proto-PNe (e.g., CRL\,2688) and very evolved bipolar PNe (e.g., 
M\,1-75), where the available flux of ionizing photons is small, the 
main excitation mechanism of the H$_2$ molecule would be shocks, 
whereas in intermediate evolutionary stages, fluorescence might dominate 
\citep{2003MNRAS.344..262D}.

The plot in Figure~\ref{plot_H2Brg}-{\it left} reveals a loose positive 
correlation between the H$_2$ 1-0 S(1)/H$_2$ 2-1 S(1) and H$_2$ 1-0 
S(1)/Br$\gamma$ line ratios.  
When the dominant excitation mechanism is taken into account 
(Figure~\ref{plot_H2Brg}-{\it right}), this plot shows that 
H$_2$-brighter sources present dominant shock excitation, 
whereas sources dominated by UV excitation typically have Br$\gamma$ 
emission brighter than H$_2$ 1-0 S(1).  
This confirms shock excitation being more efficient than UV excitation, 
and thus producing brighter H$_2$ emissions.  
We also note that, whereas the sample of H$_2$-brighter sources are mostly 
shock excited, there is a non-negligible sub-sample of Br$\gamma$-brighter 
sources that present also shock excitation.  
As a result, the trend between the H$_2$ 1-0 S(1)/H$_2$ 2-1 S(1) and 
H$_2$ 1-0 S(1)/Br$\gamma$ line ratios, which is rather sharp for the 
H$_2$-brighter soures, becomes less evident for the Br$\gamma$-brighter 
sources.

The horizontal lines in the plot mark the values of the 
H$_2$ 1-0 S(1)/H$_2$ 2-1 S(1) ratio typically assumed as 
lower limit for shock excitation ($>$10) or upper limit 
for UV excitation ($<$3). 
Within the line ratio uncertainties, the excitation of the different 
sources in this plot  
generally
follow these rules.  

%

\section{Conclusions}

We have presented near-IR $K$-band intermediate-dispersion 
spatially-resolved spectroscopic observations of a limited 
sample of bipolar PNe.  
The spectra have been used to determine the excitation mechanism 
of the H$_2$ molecule using standard line ratios diagnostics.  
Indeed, our spectroscopic observations confirm that shock-excitation is 
found to be prevalent among R-BPNe with broad equatorial rings, whereas 
W-BPNe with narrow equatorial waists present either UV excitation at 
their cores or shock-excitation at their bipolar lobes.  
The different excitation mechanisms (UV or shock) of the H$_2$ molecules 
seem to imply different levels of the H$_2$ emission with respect to that 
of Br$\gamma$, with larger H$_2$ 1-0 S(1)/Br$\gamma$ line ratios for 
shock-excited BPNe, i.e., mostly R-BPNe.

We have investigated possible correlations between different near-IR line 
ratios, H$_2$ excitation, and detailed bipolar morphology among PNe with 
available information in the literature.  
This study confirms that R-BPNe are in average brighter in H$_2$ and 
show dominant shock excitation.  
We propose the H$_2$ 1-0 S(1)/Br$\gamma$ line ratio as a preliminary 
criterion for the excitation mechanism of the H$_2$ molecules in BPNe, 
especially in evolved BPNe with a broad equatorial ring, where a high 
line ratio is very likely the result of shock-excitation.

\section*{Acknowledgments}

M.A.G.\ and R.A.M.-L.\ are supported by the Spanish MICINN (Ministerio de 
Ciencia e Innovaci\'on) grant AYA 2011-29754-C03-02 co-funded with FEDER 
funds.
R.A.M.-L.\ also acknowledges support by Mexican CONACYT (Consejo Nacional 
de Ciencia y Tecnolog\'\i a) grant No.\ 207706.
GR-L acknowledges support from CONACYT (grant 177864), CGCI, PROMEP and SEP
(Mexico).
LFM is supported by the Spanish MICINN grant AYA 2011-30228-C3-01 and 
MINECO grant AYA 2014-57369-C3-3-P, both co-funded by FEDER funds.  
Based on observations made with the Italian Telescopio Nazionale Galileo 
(TNG) operated at the Observatorio del Roque de los Muchachos, La Palma, 
Spain, by the Fundaci\'on Galileo Galilei of the INAF (Istituto Nazionale 
di Astrofisica). 
Based on observations collected at the Centro Astron\'omico Hispano Alem\'an 
(CAHA) at Calar Alto, operated jointly by the Max-Planck Institut f\"ur 
Astronomie and the Instituto de Astrof\'\i sica de Andaluc\'\i a (IAA-CSIC).

\newpage

\bsp

\label{lastpage}

\end{document}